\documentclass[12pt,preprint]{aastex}
\shorttitle{Acceleration of dusty winds in late-type stars II}
\shortauthors{Vidotto and Jatenco-Pereira}

\newcommand{\aw}{Alfv{\'e}n waves}

\begin{document}

\title{THE EFFECTS OF ALFV{\'E}N WAVES AND RADIATION PRESSURE IN DUSTY 
  WINDS OF LATE-TYPE STARS II: DUST-CYCLOTRON DAMPING}

\author{A. A. Vidotto and V. Jatenco-Pereira\altaffilmark{1}}
\affil{Instituto de Astronomia, Geof{\'i}sica e Ci{\^e}ncias
  Atmosf{\'e}ricas, Universidade de S{\~a}o Paulo, Rua do Mat{\~a}o
  1226, 05508-900 S{\~a}o 
  Paulo, SP, Brazil} 

\begin{abstract}
There are in the literature several theories to explain the mass loss
in stellar winds. In particular, for late-type stars, some authors
have proposed a wind model driven by an outward-directed flux of
damped Alfv{\'e}n waves. The winds of these stars present great
amounts of dust particles that, if charged, can give rise to new wave
modes or modify the pre-existing ones. In this work, we study how the
dust can affect the propagation of Alfv{\'e}n waves in these winds
taking into account a specific damping mechanism, dust-cyclotron
damping. This damping affects the Alfv{\'e}n wave propagation near the
dust-cyclotron frequency. Hence, if we assume a dust size distribution,
the damping occurs over a broad band of wave frequencies. In this work,
we present a model of Alfv{\'e}n wave-driven winds using the
dust-cyclotron damping mechanism. On the basis of coronal holes in the Sun,
which present a superradial expansion, our model also assumes a
diverging geometry for the magnetic field. Thus, the mass, momentum,
and energy equations are obtained and then solved in a self-consistent
approach. Our results of wind velocity and temperature profiles for a
typical K5 supergiant star shows compatibility with observations. We
also show that, considering the presence of charged dust particles,
the wave flux is less damped due to the dust-cyclotron damping than
it would be if we consider some other damping mechanisms studied in
the literature, 
such as nonlinear damping, resonant surface damping, and turbulent damping.
\end{abstract}


\section{INTRODUCTION}
Cool supergiant stars present high mass loss rates, $\dot{M} \sim
10^{-8} - 10^{-5} {\rm ~M}_\odot {\rm ~yr}^{-1}$, and low
terminal velocities, $u_\infty \sim 10 - 100{\rm ~km~s}^{-1}$
\citep{1979ARA&A..17..275C}, which are less than the escape
velocity at the stellar surface ($v_{e_{0}}$). Typically, it is
observed that $u_\infty \lesssim v_{e_{0}}/2$. Although the physical
mechanism that drives these winds is still poorly known, driving the wind
with magnetohydrodynamic waves is one of the most promising mechanism
proposed to date. Several authors \citep[e.g.,][]{2001A&A...377..522L,
2001MNRAS.327..403E, 2003A&A...399..589H, 2005A&A...433.1101W}, motivated by 
the observed high stellar luminosity and low effective temperature of
these stars, have
proposed radiatively dust driven wind models to explain the wind
acceleration. Acoustic waves generated at a pulsating phase could
provide a propitious ambient medium for dust formation; this is, a high-density
region with low temperature. Hence, radiative pressure acting on dust
particles could accelerate the grain and, if the gas and the dust were
dynamically coupled, grains could drag the gas and provide the mass
loss. However, this mechanism fails when gas and dust are not well
coupled \citep{2003A&A...404..789S}. Another failing aspect of this
mechanism is that there are observations attesting that the wind forms
before the grain formation point \citep{1995ApJ...444..424C}. 

In this aspect, Alfv{\'e}n waves might play an important role in these
scenarios, producing significant mass loss and low terminal velocities
\citep[][p. 294]{lamerscassinelli}. In this mechanism, the wave
magnetic pressure gradient is added into the momentum equation in
order to drive the wind.  

\citet{1980ApJ...242..260H} investigated the effects of Alfv{\'e}n
waves on the outer atmospheres of cool supergiant stars. They
estimated an initial wave flux of $\sim 10^6 {\rm ~ergs~cm}^{-2}{\rm
~s}^{-1}$. This flux is of the same order of magnitude as the one
estimated for the solar wind \citep{1998A&A...339..208B}. However,
they found that undamped waves led to very high terminal wind
velocities. In order to obtain terminal velocities that were in
reasonable agreement with observations of 
cool supergiants stars, it was necessary to dissipate the waves
on a length scale of $\sim r_0$ ($r_0$ is the star radius). The physical
mechanism responsible for the damping was left unspecified. 
\citet{1983ApJ...275..808H} argued that this  length scale was very
restrictive and that we could not observe a large variety of stars
if we had such selective dissipation scale for the waves. This problem
was known as the \emph{fine tuning problem}. \citet{1989A&A...209..327J} 
showed that flux tubes expanding superradially from the stellar surface, as is
observed to happen in solar coronal holes \citep{2005ApJS..156..265C},
could account for a broad band of dissipation scales, overcoming this
problem.

In \citet[hereafter Paper I]{2002ApJ...576..976F}, a hybrid model was
presented, in which the effects of a flux of \aw\ acting together with
radiation pressure on grains were analyzed as acceleration mechanisms
of winds of cool supergiant stars. In that model, an empirical
temperature profile was introduced. The presence of grains had two
roles in that model: grains were used to simulate a strong damping
mechanism for the waves and as an additional acceleration mechanism due to
the exertion of radiation pressure on them. Hence, a more realistic
wind model showing results consistent with observations was obtained.

Here, we present a model in which a flux of \aw\ propagating outward
from the star is damped due to the interaction between the waves and
charged dust particles, thus, accelerating the wind. In order to obtain both
the temperature and the velocity profiles, we solve the mass, the
momentum, and the energy equations of the wind. In this work, we
improve the model presented in Paper I by: (1) inserting the energy
equation into the model in order to obtain the temperature profile of
the wind; and (2) inserting a more realistic damping mechanism for
the waves due to the presence of grains. Here, we examine the
influence of this damping mechanism instead of simulating a strong
damping. In \S\ref{sec.model}, we describe the model used to
accelerate the wind: we describe the modified wave dispersion relation
due to the presence of charged dust in \S\ref{subsec.damping}, the
geometry considered for the magnetic field lines in
\S\ref{subsec.geometry}, the equations that describe the wind dynamics
in \S\ref{subsec.equations}, and the radiation pressure on dust in
\S\ref{subsec.radiation}. In \S\ref{sec.results} and \S\ref{sec.conc},
we present our results and conclusions.

\section{MODEL}\label{sec.model}
\subsection{Dust-Cyclotron Damping} \label{subsec.damping}
The propagation and damping of Alfv{\'e}n waves in a dusty plasma has
been considered by many authors \citep{1987ApJ...314..341P, mr92,
1992PhyS...45..504S, 2003ApJ...597..970F}. Dust grains immersed in an
ambient plasma become charged due to plasma ion and electron fluxes
into grain surfaces. Although the number of dust particles is smaller
than the number of ions, the process of dust charging is efficient and
these particles can obtain charges $q_d = -z_d e$ for $z_d$ on the
order of $\sim 10^0 - 10^3$ in astrophysical media \citep{g89,mr94}. Once
charged, these particles suffer the influence of the magnetic field
which gives rise to a cyclotron frequency. Thus, they can modify the plasma
behavior in different ways. In particular, charged dust particles
introduce a cutoff (resonance) in the Alfv{\'e}n wave spectrum at the
dust cyclotron frequency. For the ions, this resonance occurs in a narrow
range of higher frequencies, which are unimportant in the systems under
consideration. On the other hand, the dust cyclotron resonance occurs
at low frequencies and it can be an important damping mechanism for
the waves. 

If a distribution of grain sizes is considered, then we obtain a band
of resonance frequencies from the minimum dust-cyclotron
frequency $\omega_{\rm min}$ to the maximum dust-cyclotron frequency
$\omega_{\rm max}$. According to \citet{mrn77}, we can describe the
distribution of grain sizes in the interstellar medium by a power law
(MRN distribution):
\begin{equation}\label{eq.mrn} 
f(\mathcal{R})\, d\mathcal{R} = c_p \, \mathcal{R}^{-p} \,
d\mathcal{R} \, ,
\end{equation}
where $c_p = ({p-1})/({1 - a_m^{1-p}})$, $\mathcal{R} = a/a_{\rm min}$ is a
dimensionless radius, $a$ is the grain radius, $a_m = a_{\rm
  max}/a_{\rm min}$, and $a_{\rm min}$ and 
$a_{\rm max}$ are the minimum and the maximum dust radii,
respectively. Observationally, the parameter $p$ depends on the dust  
constituent and on the environment. In this work we consider $p =
4$, as used by \citet{2002PhPl....9.4845C}. In this case, we have $c_4
= {3}/({1 - a_m^{-3}})$. 

The wave propagation in dusty plasma is modified and some new and
interesting effects take place \citep{wn99,cramer2001}. The dispersion
relation of the Alfv{\'e}n waves, with frequencies smaller than the
ion cyclotron frequency, considering dust particles with constant
charges in a neutral and cold dusty plasma, is given by
\citep{2002PhPl....9.4845C}
\begin{equation}\label{k}
k^2_z = u_1 \pm u_2\, ,
\end{equation}
where:
\begin{equation}
u_1 = \frac{\omega^2 \Omega_{i}^2}{v_{Ai}^2 ( \Omega_{i}^2-
  \omega^2)} + \frac{\omega^2 \Omega_{d0}^2}{s v_{Ad}^2} \int_1^{a_m}
\frac{f(\mathcal{R}) d\mathcal{R}}{\mathcal{R}
  (\Omega_{d0}^2/\mathcal{R}^4 - \omega^2)} \, ,
\end{equation}
and
\begin{equation}
u_2 = \frac{\omega^3 \Omega_{i}}{v_{Ai}^2 ( \Omega_{i}^2- \omega^2)}
+ \frac{\omega^3 \Omega_{d0}}{s v_{Ad}^2} \int_1^{a_m}
\frac{\mathcal{R} f(\mathcal{R}) d
  \mathcal{R}}{\Omega_{d0}^2/\mathcal{R}^4 - \omega^2} \, .
\end{equation}
Here $s = c_4 \ln (a_m)$, $\omega$ is the angular wave
frequency, $\Omega_{i}$ is the ion cyclotron frequency, $\Omega_{d0} =
q_{\rm min} B / (m_{\rm min} c) \equiv \omega_{\rm max}$ is the
maximum dust cyclotron frequency (i.e., for the minimum dust grain
with mass $m_{\rm min}$ and charge $q_{\rm min}$), and $v_{A \alpha}=
{B}/({\sqrt{4 \pi \rho_\alpha}})$ is the Alfv{\'e}n speed of the
$\alpha$ species. Here $\alpha = i$ and $d$ for ions and dust,
respectively, $B$ is the magnetic field intensity, $\rho_\alpha$ is
the $\alpha$ species density, and $c$ is the speed of light. In order
to calculate the grain mass, we assume spherical silicates grains with
$\rho_{\rm sil} = 3.3$~g cm$^{-3}$.

Dust grains are usually negatively charged, with many electrons
collected on each particulate. The mean grain particle charge can be
obtained for each dust radius by considering the charge current equilibrium
over the dust surface. The equilibrium equation, if we consider a Maxwellian
distribution of velocities, is given by \citep{1997Ap&SS.256...85V}
\begin{equation}\label{eq.charge}
\frac{\omega_{p_{i}}^2}{\omega_{p_{e}}^2}   \left( 1 + \frac{z_d
  e^2}{ a k_B T} \right)  = \frac{v_{T_{i}}}{v_{T_{e}}} \exp
\left\{  - \frac{z_d e^2}{ a k_B T} \right\} \, .
\end{equation}
The plasma frequency, $\omega_{p_\beta}$, for electrons ($\beta =
e$) or ions ($\beta = i$) is given by:
\begin{equation}
\omega_{p_\beta} = \sqrt{\frac{4 \pi \rho_\beta q_\beta^2}{m_\beta^2}}
\end{equation}
and the thermal velocity, $v_{T_{\beta}}$, is given by:
\begin{equation}
v_{T_{\beta}} = \sqrt \frac{k_B T}{m_\beta} \, ,
\end{equation}
where $k_B$ is the Boltzmann constant. 
%Here, we assumed that electrons and ions have the same temperature $T$. 

The left-hand polarized wave (minus sign in eq.~[\ref{k}])
interacts with ions and is not affected by the negatively charged dust
particles resonance. The right-hand polarized wave (plus sign in
eq.~[\ref{k}]) is the mode damped by the dust resonance. In the case
of $\omega_{\rm min} < \omega < \omega_{\rm max}$, the integral over the
particle's radius has singularities, whose residues give the complex
part of the wave number ($k_z \equiv k_\mathcal{R} + i k_\mathcal{C}$)
and that leads to the dust-cyclotron damping of the waves, with
damping length:
\begin{equation}\label{L}
L(\omega) = \frac{2 \pi}{k_\mathcal{C}(\omega)}  \, .
\end{equation}

We adopt a particle radius distribution from $a_{\rm min} = 5.0 \times
10^{-7}$~cm to $a_{\rm max} = 2.5 \times 10^{-5}$~cm
\citep{1991ApJ...382..606R}. In order to calculate $\rho_d$, we adopt
a gas-to-dust ratio given by:
\begin{equation}
R_{gd} = \frac{\rho}{\rho_d}\, ,
\end{equation}
where $\rho$ is the gas density. 

In this work, we do not develop a self-consistent model for the grain
nucleation. According to the 
model of \citet{1989A&A...223..227D} for dust-driven winds around carbonated
stars, near the critical point ($\sim 1.5~r_0$) there is an intense
grain formation. For $r < 1.5~r_0$, grains are only formed locally, and
beyond $2.0~r_0$, dust is already condensed into an envelope around
the star. On the basis of this scenario, we adopt the following
situation. For $r \gtrsim 2.0~r_0$, $R_{gd} \simeq 200$
\citep{1985ApJ...293..273K, 2005A&A...439..171H}, which implies that the
dust envelope has already been formed. Near the wind base, $R_{gd} \gg
200$,  which implies the existence of a negligible dust quantity.  Between
the wind base and $\sim 2.0~r_0$, $R_{gd}$ decreases exponentially,
which indicates that the grains are being formed until $R_{gd}$ reaches its
minimum value of $\simeq 200$.

\subsection{Magnetic Field Geometry}\label{subsec.geometry}
It is beyond the scope of this paper to discuss magnetic field generation
in late-type stars. However, many works studying field generation have been
presented in the literature \citep{2001Natur.409..485B,
2002MNRAS.329..204S, 2004A&A...423.1101D}. These works are motivated
by observations of, for instance, maser polarization
\citep{2002A&A...394..589V} or X-ray emission
\citep{2003ApJ...598..610A} from late-type stars, which are taken as
an indicative of the presence of magnetic fields around them. Hence,
jointly with perturbations (e.g., convective motions), the basic
ingredient for Alfv{\'e}n wave generation is present in the
atmospheres of late-type stars.  

\begin{figure}
\epsscale{.7}
\plotone{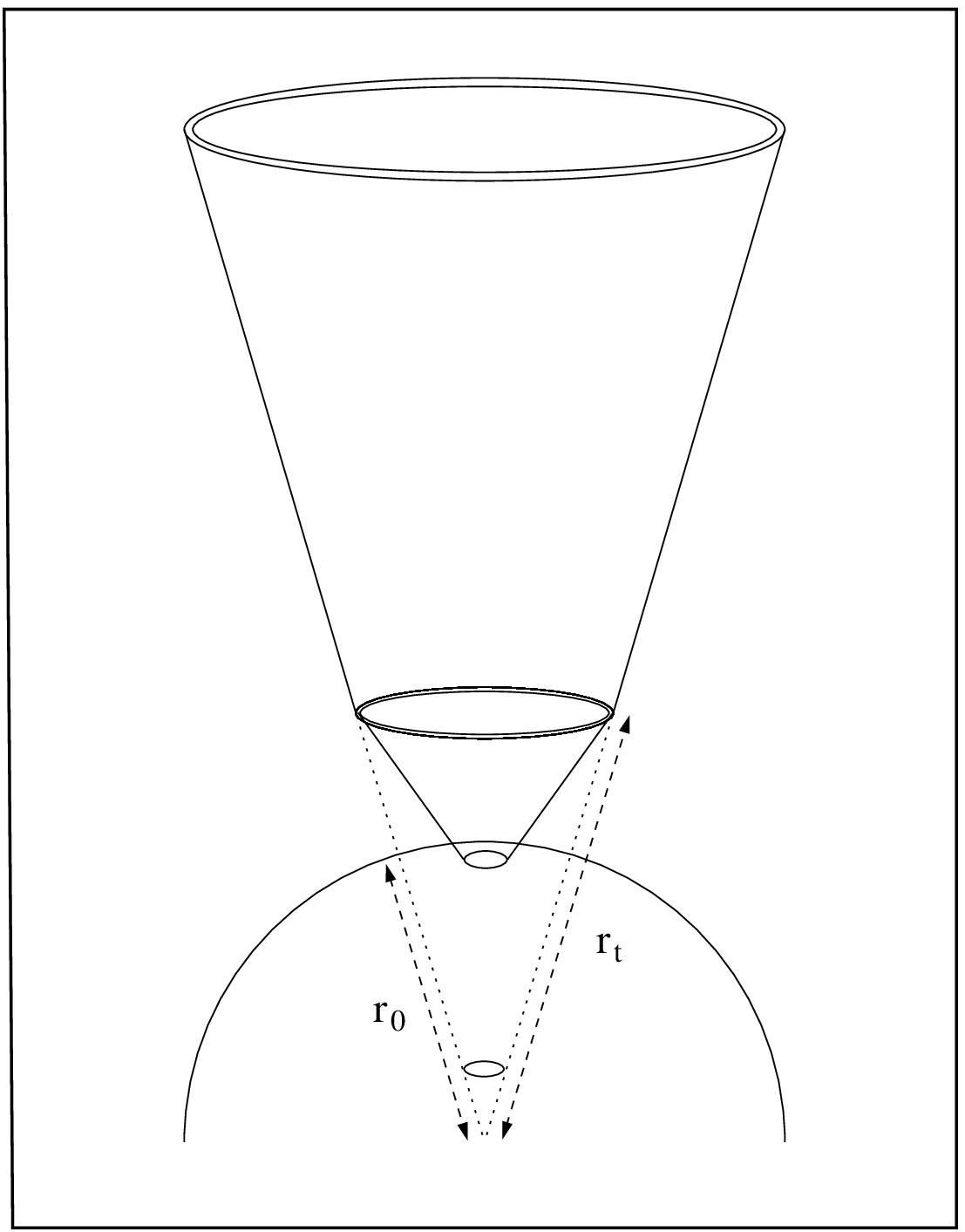}
\caption{Geometry used in order to reproduce coronal holes (not to
  scale).\label{geometry}} 
\end{figure}

On the basis of this assumption, \citet{1989A&A...209..327J} suggested a
model for mass loss in late-type stars driven by an outward flux of
damped Alfv{\'e}n waves that took into account our knowledge of coronal
holes in the Sun. It is known that the coronal holes are the source of
the high-speed solar wind streams at the Earth's orbit
\citep[e.g.,][]{1999ApJ...522.1148P}. The role of Alfv{\'e}n waves in
accelerating the fast solar wind and the heating of plasma in the open
magnetic structures of the solar corona has been discussed since the
pioneering work of \citet{parker} \citep[e.g.,][among 
others]{rosenthal02, suzuki05}. The area expansion of a solar 
coronal hole is $5-13$ times greater than that for a radial expansion
\citep{2002ApJ...566..562L, 2005ApJ...629L..61E,
2005Sci...308..519T}. For simplicity, we assume that in the wind of
a typical K5 supergiant star, the non-radial expansion factor is  
\begin{equation}
F = \frac{\Omega}{\Omega_0} = 10 \, ,
\end{equation}
where $\Omega_0$ and $\Omega$ are the solid angles at the stellar surface
and at the edge of the coronal hole, respectively. Hence, in order to
reproduce the coronal holes, we use a diverging geometry 
suggested by \citet{1982A&A...114..303K} for the magnetic field. The
cross section of a flux tube at a radial distance $r$ is given by
\begin{displaymath}
A(r) = \left\{ \begin{array}{ll}
A(r_0) (r/r_0)^S & {\rm ~if~ } r \leq r_t \\
A(r_0) (r_t/r_0)^S (r/r_t)^2 & {\rm ~if~ } r > r_t \, ,
\end{array} \right.
\end{displaymath}
where $S$ is a parameter that determines the divergence of the
geometry up to the transition radius $r_t$, as shown in
Figure~\ref{geometry}. For a given $S$, $r_t$ is easily obtained: 
\begin{equation}
F = \frac{\Omega}{\Omega_0} = \frac{A(r_t)/r_t^2}{A(r_0)/r_0^2} =
\left( \frac{r_t}{r_0} \right)^{S-2} \, . 
\end{equation}
Hence, we have
\begin{equation}
r_t = F^{1/(S-2)} \, r_0 \, .
\end{equation}
Then, when we consider this geometry, conservation of magnetic flux yields
a magnetic field intensity $B(r) \propto A(r) ^{-1}$.

\subsection{The Wind Equations}\label{subsec.equations}
The mass, momentum, and energy equations are the equations responsible
in driving the wind:

\paragraph{The mass equation} It expresses the continuity of
matter. For a flow velocity $u$, we have:
\begin{equation}\label{mass}
\rho u A(r) = \mathcal{C} \, .
\end{equation}
The constant $\mathcal{C}$ in equation (\ref{mass}) is computed at
the stellar surface: $\mathcal{C} = \rho_0 u_0 A(r_0)$, where the subscript
``$0$'' indicates that the parameter is being evaluated at $r = r_0$.

\paragraph{The momentum equation} In this model, we assume that the
forces acting on the wind particles are 
the gravitational force, the gas pressure gradient, and the wave
magnetic pressure gradient. Hence, if we assume a steady flow, the
equation of motion can then be written as: 
\begin{equation}\label{momentum}
\rho u \frac{du}{dr} = - \rho \frac{G M_\star}{r^2} - \frac{dP}{dr} -
\frac{1}{2} \frac{d{\epsilon}}{dr}   \, , 
\end{equation}
where $P= \rho k_B T / m$ is the gas pressure, $m$ is the average mass
per particle, $\epsilon$ is the energy density of the Alfv{\'e}n waves,
and $G M_\star /r^2$ is the gravitational acceleration. The last term
in equation (\ref{momentum}) is the wave magnetic pressure
gradient. As ${dP}/{dr}$ and ${d{\epsilon}}/{dr}$ are both negative
quantities, we can see that the gas and the wave pressure gradients
act in the opposite direction to the gravitational attraction. 

\paragraph{The energy equation} In order to compute the temperature
profile of the wind, one needs 
to evaluate the energy equation. In the present model, the wind
temperature is determined through the balance between wave 
heating, adiabatic expansion, and radiative cooling. If we neglect
conduction and assume a perfect gas, we write the energy equation as
\citep{1982ApJ...261..279H}
\begin{equation} \label{energy}
\rho u \frac{d}{dr} \left( \frac{u^2}{2} + \frac52 \frac{k_B T}{m} -
\frac{v_e^2}{2} \right) + \frac{u}{2} \frac{d \epsilon}{dr} = (Q
- P_R) \, , 
\end{equation}
where $v_e$ is the escape velocity. The term $({u}/{2}) {d
\epsilon}/{dr}$ is the rate at which the waves do work on the gas. The
quantity $Q$
is the wave heating rate, i.e., the rate at which the gas is being
heated due to the dissipation of wave energy, and $P_R$ is the
radiative cooling rate, both in units of ergs~cm$^{-3}$~s$^{-1}$. 

The radiative cooling rate is given by: 
\begin{equation} 
P_R = \Lambda \, n_e \, n_H \, ,
\end{equation}
where $n_e$ is the electron density, $n_H$ is the hydrogen density, and
$\Lambda$ is the radiative loss function. Here, we adopt the value of $\Lambda$
given by \citet{1993A&A...273..318S} and calculate $n_e$ as did
\citet{1980ApJ...242..260H}. 

In order to evaluate the wave heating rate $Q$, one needs to know how
the wave energy is being dissipated. In our model, we assume that
Alfv{\'e}n waves are damped by the dust-cyclotron damping mechanism,
which has a damping length $L(\omega)$ given by equation~(\ref{L}). It
is importante to note that each wave frequency is damped by a
different amount (see~\S\ref{sec.results} for a discussion). Hence,
the wave heating rate must be different for each wave frequency. The
wave heating rate that enters equation (\ref{energy}) is, then,
calculated as follows: 
\begin{equation}
Q =  \int_{\omega_{\rm min}}^{\omega_{\rm max}} Q(\omega) \, d \omega  \, ,
\end{equation}
where $Q(\omega)$ is the wave heating rate per frequency and it is
directly related to the wave energy density per unit frequency $\epsilon
(\omega)$ and the damping length $L(\omega)$ associated with it. Hence,
we have \citep{1973ApJ...181..547H}
\begin{equation}
 Q(\omega) =\frac{\epsilon (\omega)}{L(\omega)} (u + v_A) \, .
\end{equation}

Similarly, the wave energy density $\epsilon$ needs to be calculated
over the frequency band. Hence, we have: 
\begin{equation}
\epsilon = \int_{\omega_{\rm min}}^{\omega_{\rm max}} \epsilon
(\omega)\,  d \omega 
\end{equation}
and, for each frequency, it is calculated as
\citep{1989A&A...209..327J}:
\begin{equation} \label{eq.epsilon}
 \epsilon (\omega) = \epsilon_0 (\omega) \frac{M_0}{M} \left
 (\frac{1+M_0}{1+M} \right )^{2} \exp{ \int_{r_0}^r- {\frac{1}{L
 (\omega)}} \, dr' } \, , 
\end{equation}
where $M=u/v_A$ is the Alfv{\'e}n Mach number with \mbox{$v_A = (B /
\sqrt{4 \pi \rho})$}. 

\citet{1989JGR....9411739T} observed an Alfv{\'e}n wave spectrum
propagating in the solar wind and inferred for the low-frequency
range that $\epsilon (\omega) \propto \omega^{-\beta}$, with $\beta
\sim 0.6$. Here, we assume that the initial energy density of \aw\
propagating in the winds of cool supergiant stars behaves similarly:
\begin{equation}\label{eq.spec}
\epsilon_0 (\omega) = \epsilon_0 (\omega_0) \left(
\frac{\omega}{\omega_0} \right) ^{-\beta} \, ,
\end{equation}
where $\epsilon_0 (\omega_0)$ is the initial wave energy density at any
frequency $\omega_0$. The initial energy density and the initial wave
flux evaluated at $\omega$ are related by \citep{1989A&A...209..327J}: 
\begin{equation}
\phi_{A_{0}} (\omega) = \epsilon_0 (\omega) v_{A_{0}} \left( 1 +
\frac32 M_{0} \right) \, .
\end{equation}
In our model, we adopt an initial wave flux as estimated by
\citet{1980ApJ...242..260H}:
\begin{equation}
\phi_{A_{0}} = \int_{\omega_{\rm min}}^{\omega_{\rm max}} \phi_{A_{0}}
(\omega) \, d \omega \simeq 10^6 {\rm ~erg~cm}^{-2}~{\rm s}^{-1} \, . 
\end{equation}

From equations~(\ref{momentum}), (\ref{energy}), and
(\ref{eq.epsilon}) we can write the temperature variation as:
\begin{equation} \label{energy2} 
\frac{dT}{dr} = \frac23 \frac{T}{r} \left[ \frac{r(Q-P_R)}{\rho u (k_B
    T /m)} - \left( Z + \frac{r}{u} \frac{du}{dr} \right) \right]
\end{equation}
and the velocity variation as:
\begin{eqnarray}\label{momentum2}
\frac{1}{u} \frac{du}{dr} \left[ u^2 - \frac53 \frac{k_B T}{m} -
\frac{\epsilon}{4 \rho} \left( \frac{1+3M}{1+M}
\right)   \right]  &=& \nonumber \\
= \frac{Z}{r} \left[ - \frac23
\frac{r(Q-P_R)}{Z \rho u} + \frac{\epsilon}{4 \rho}
\left( \frac{1+3M}{1+M} \right)  \right. &+& \nonumber \\
\left. + \frac53 \frac{k_B T}{m} + \frac{r}{2 Z \rho} \int_{\omega_{\rm
    min}}^{\omega_{\rm max}} \frac{\epsilon (\omega)}{L(\omega)}
\, d \omega - \frac{v_e^2}{Z} \right]  \, ,  &&
\end{eqnarray}
where we define $Z$ as:
\begin{displaymath}
Z \equiv \left\{ \begin{array}{ll}
S & {\rm ~if~} r \leq r_t \\
2 & {\rm ~if~} r > r_t \, .
\end{array} \right.
\end{displaymath}
Hence, in order to obtain the velocity and the temperature profiles of the
wind, we then solve equations~(\ref{mass}), (\ref{energy2}), and
(\ref{momentum2}), simultaneously, for the wind of a typical K5
supergiant star.

\subsection{Radiation Pressure}\label{subsec.radiation}
As in Paper I, we include the radiation pressure into
the wind equation using an effective escape velocity:
\begin{equation}
\frac12 v_e^2 = \frac{G M_\star}{r} (1-\Gamma) \, ,
\end{equation}
where $\Gamma$ is the ratio between the force per unit volume exerted
on $n_d$ grains of radius $a$ situated at a distance $r$ from a star
of luminosity $L_\star$ [$n_d \pi a^2 Q_d L_\star/(4 \pi c r^2)$] and
the gravitational force per unit volume ($\rho G M_\star/r^2$):
\begin{equation}
\Gamma = \frac{n_d}{\rho} \frac{\pi a^2 Q_d L_\star}{4 \pi c G
  M_\star}\, .
\end{equation}
Here $Q_d$ is the mean extinction efficiency factor of the grain. In order
to analyze how effective the radiation pressure is on the grains, we
make a simple estimate. For silicate grains with $a \simeq 5 \times 
10^{-6}$~cm in the atmosphere of a star with an effective temperature
of $3500$~K, we have $Q_d \sim 10^{-3}$
\citep{1974ApJS...28..397G}. We can write: $n_d/\rho = \rho_d/(\rho
m_d) = (\mathcal{R}_{gd} m_d)^{-1}$. Then, if we assume that $\mathcal{R}_{gd} 
\simeq 200$, we obtain $\Gamma \simeq 0.03$; that is, the radiation pressure
modifies the square of the escape velocity by only $\sim 3\%$. This is
not a significant modification. 

In Paper I, larger grains were considered. As $Q_d$ is more important
for large grains, it was shown in Paper I that radiation
pressure leads to significant higher terminal velocities, although it
can not initiate the wind. It is observed that the wind forms before
the grain formation point \citep{1995ApJ...444..424C}, which indicates that
another mechanism is taking place in order to drive the wind.

In the present model, we observe only a negligible difference in the
terminal velocity if we consider the radiation pressure or not. This
happens because we are adopting the radius distribution given by
equation~(\ref{eq.mrn}), which causes our model to have many more
small grains than large ones.

\section{RESULTS AND DISCUSSION}\label{sec.results}
For the solution of the set of equations~(\ref{mass}), (\ref{energy2}),
and (\ref{momentum2}), we adopted the set of initial conditions
presented in Table~\ref{parameters} for the wind of a typical K5
supergiant star. We solve these equations until $r = 300~r_0$, at
which point
the wind has already reached its terminal velocity. Some parameters
adopted here, such as the stellar mass, the stellar radius, the
initial density, and the initial magnetic field, all have of the same
values as the ones adopted by \citet{1980ApJ...242..260H} for their
model 6. The parameters calculated for the dust grains are
shown in Table~\ref{parameters_grains}. 

The momentum equation (eq.~[\ref{momentum2}]) of the wind has a
critical point when both the numerator and denominator go to zero. The
requirement that the wind velocity should increase through the
critical point determines the initial velocity at the base of the
wind, i.e., ${du} {(r=r_0)} / {dr} \geq 0$. Hence,
to find the critical solution for a given set of initial conditions,
we iterate the initial velocity $u_0$ until the solution passes through the
critical point. Together with the initial density, the initial
velocity determines the mass-loss rate.   

\begin{figure}
\epsscale{1}
\plotone{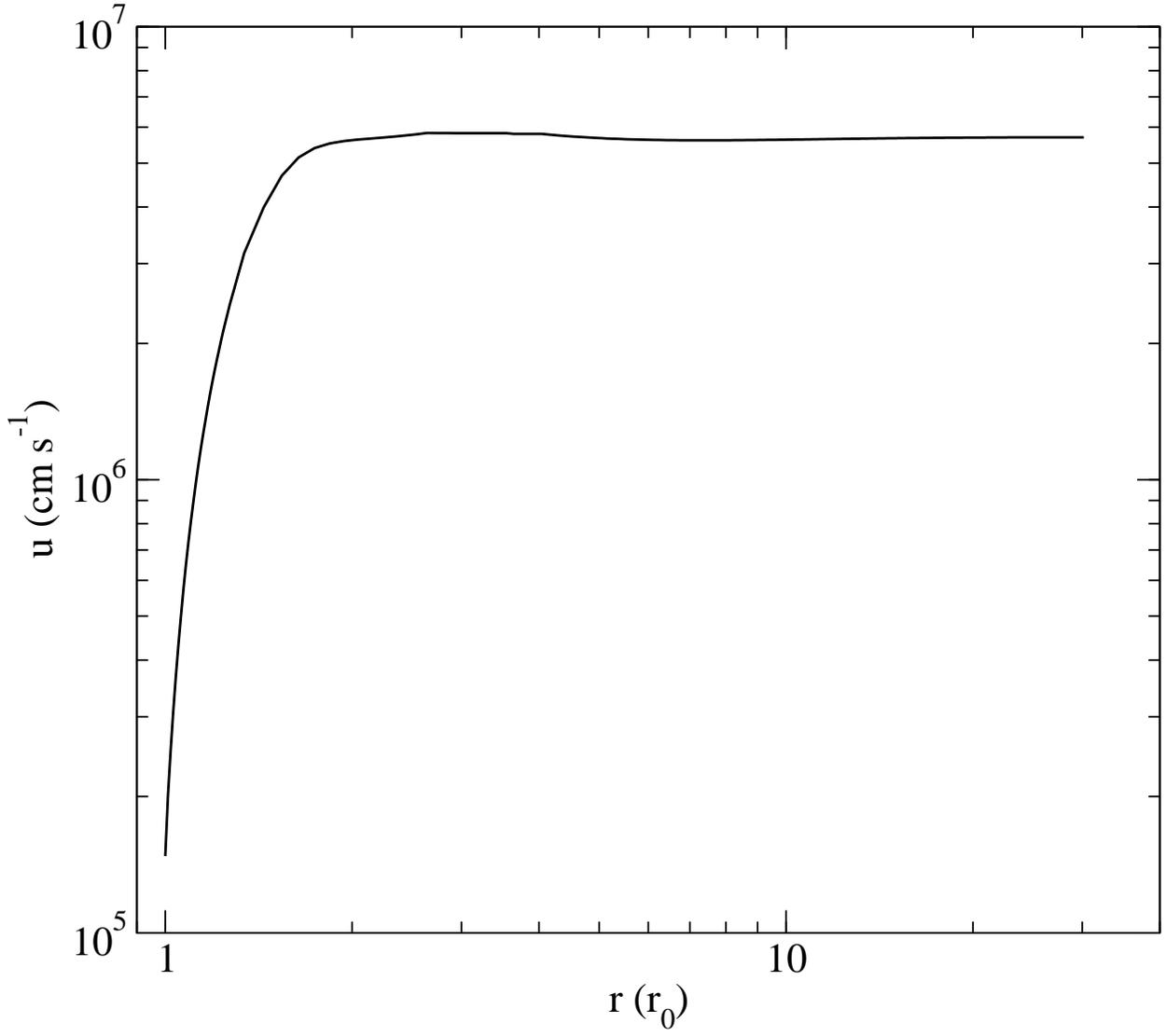}
\caption{Wind velocity profile obtained for a typical K5 supergiant
  star. \label{velocidade1}} 
\end{figure}

\begin{figure}
\epsscale{1}
\plotone{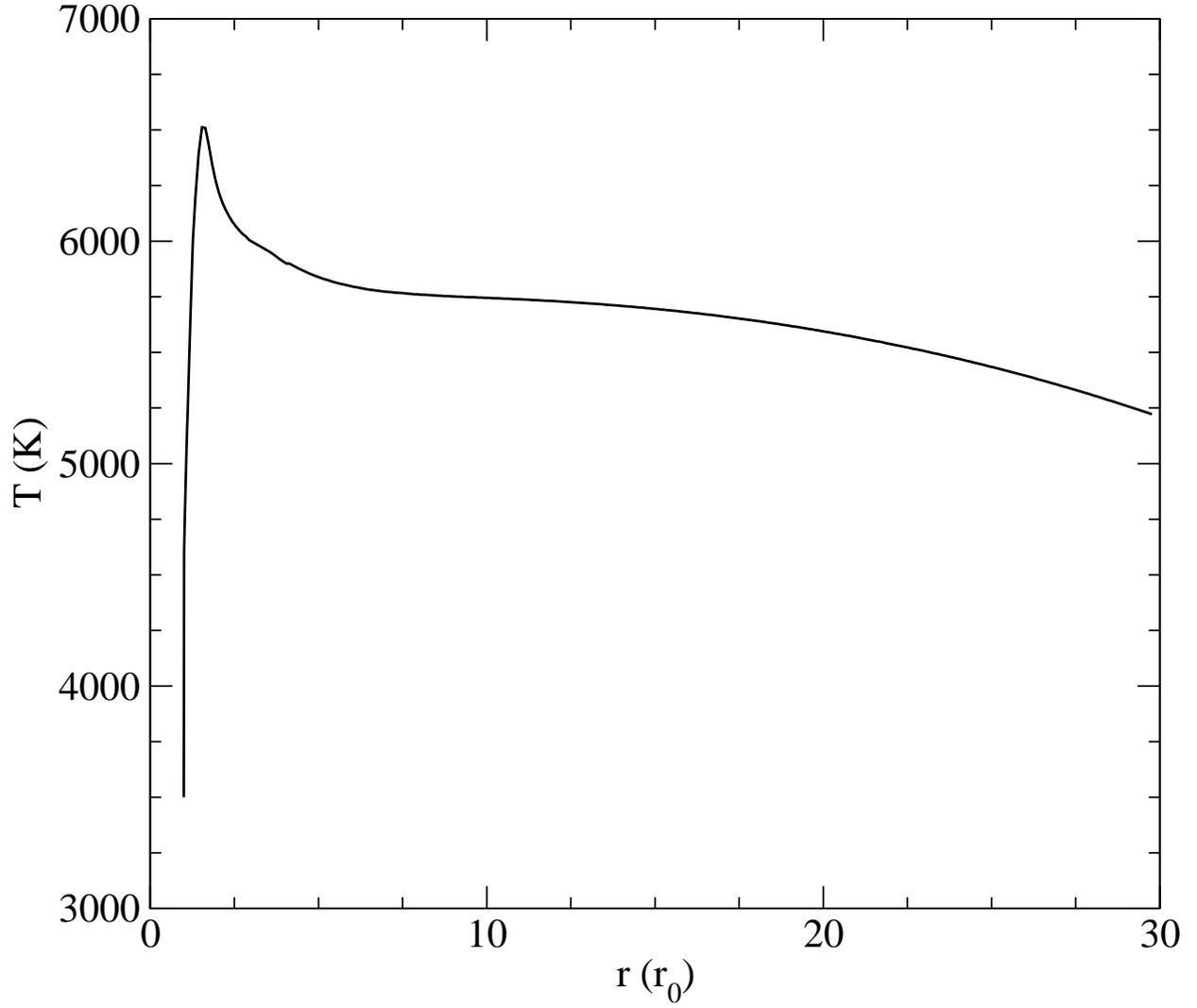}
\caption{Wind temperature profile obtained for a typical K5 supergiant
  star. \label{temperatura1}} 
\end{figure}

To simplify the problem, we calculate $ \omega_{\rm min}$ and $
\omega_{\rm max}$ at $r = r_0$ and assume that these parameters are constant
throughout the wind. The grains charges are evaluated according to
equation~(\ref{eq.charge}). Our results are shown in
Figures~\ref{velocidade1} and \ref{temperatura1}, where we plot the velocity
and the temperature profiles that we obtained. In order to
have a better view of what is happening near the stellar surface, we
plot the profiles until $r = 30~r_0$.

The results of the present model compares favorably with
observations. We expect the terminal velocity to be lower than the
surface escape velocity. From observations, we expect $u_\infty \sim
v_{e_{0}}/2$. Our model provides $u_\infty \simeq 57 $
\mbox{km s$^{-1}$} and \mbox{$\dot{M} \simeq 2.5 \times 10^{-7}$
  M$_\odot$ yr$^{-1}$}. In Figure~\ref{velocidade1}, we can see that
the wind is accelerated near the surface, even before the grain
formation point ($\sim 2~r_0$). 

As we know from solar corona observations, we expect a rise in the
wind temperature near the stellar surface. From
Figure~\ref{temperatura1}, it can be seen that the rise in the
temperature occurs very abruptly near the wind base. This heating is
explained by the mechanical dissipation of Alfv{\'e}n waves. At
larger distances ($r \gtrsim 30~r_0$), the flow expands adiabatically, hence,
in the absence of strong heating, the temperature tends to fall
monotonically with the adiabatic exponent $4/3$.

\begin{figure}
\epsscale{1}
\plotone{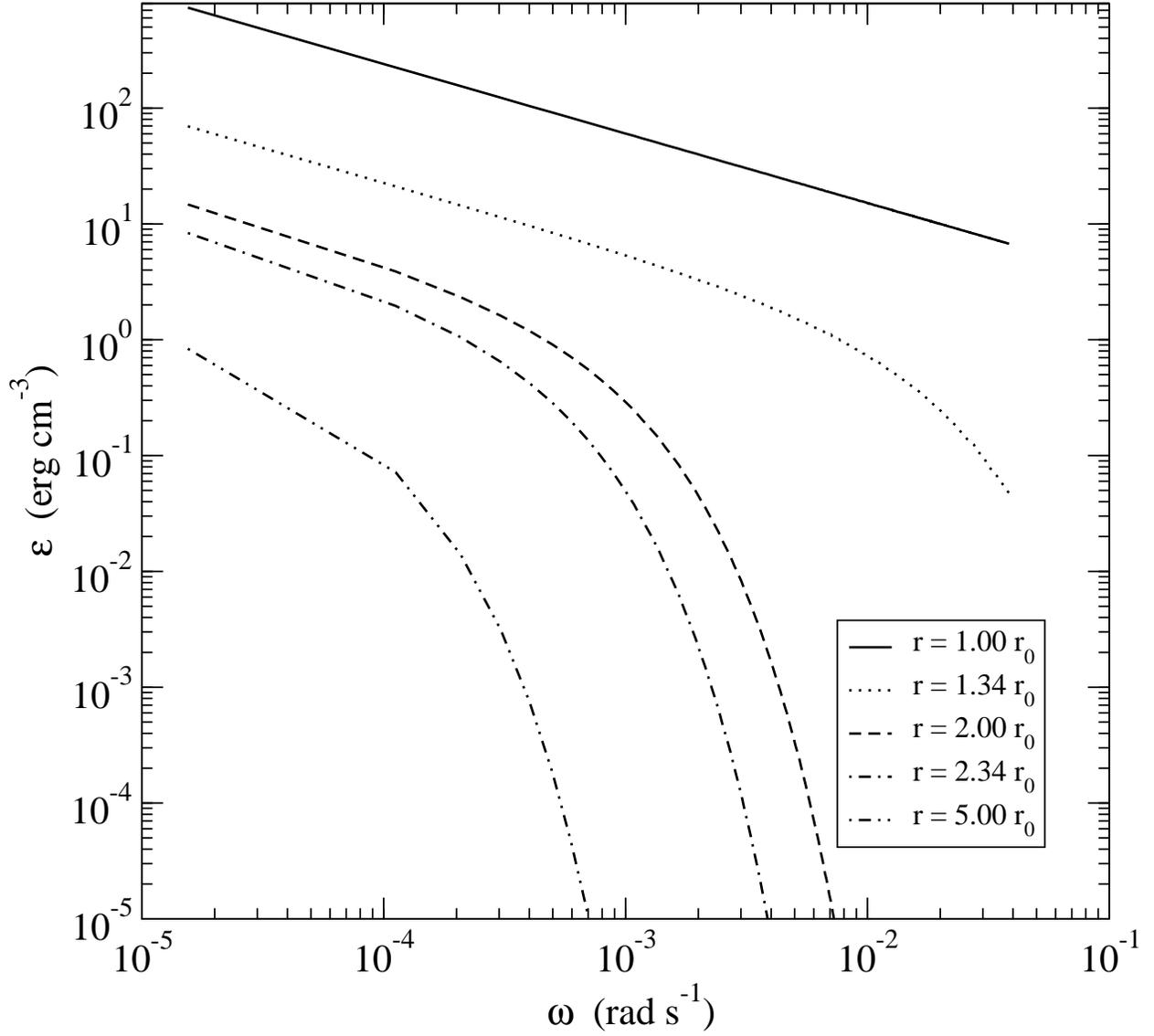}
\caption{Energy density of Alfv{\'e}n waves at different
  positions: at $1~r_0$ (solid curve), at $1.34~r_0$ (dotted curve), at
  $2~r_0$ (dashed curve), at $2.34~r_0$ (dash-dotted curve), and at
  $5~r_0$ (dash-double-dotted curve). \label{energia}} 
\end{figure}

In order to analyze the wave dissipation, we plot in
Figure~\ref{energia} the energy density of the waves at different
distances from the star. At $r = r_0$, we have a power-law spectrum
(eq.~[\ref{eq.spec}]). As we get farther from the wind base, the energy
density decreases due to the flow expansion and wave damping. In
Figure~\ref{energia}, we note that this decline is larger 
for larger wave frequencies, indicating a more efficient damping for
waves with higher frequencies. This is due to the fact that we have a
larger number of small-radius grains (see eq.~[\ref{eq.mrn}]). As the smaller
grains have larger cyclotron frequencies, they interact with the
higher frequency waves, causing a larger damping at this spectral range. 

\begin{figure}
\epsscale{1}
\plotone{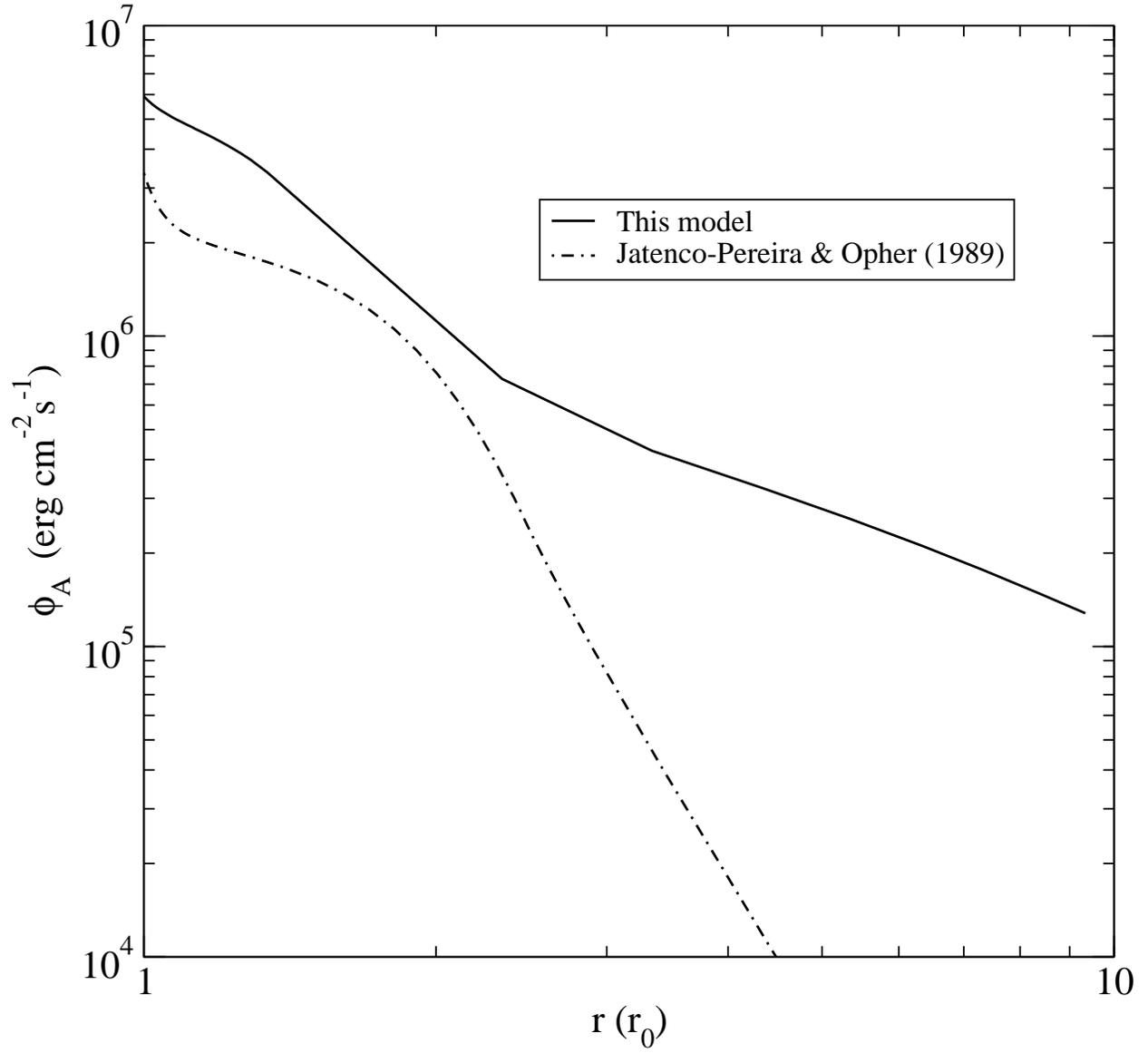}
\caption{Comparison between Alfv{\'e}n waves fluxes for this model
  (solid curve) and for the model of \citet{1989A&A...209..327J}
  (dash-dotted curve). \label{fig.fluxo}}
\end{figure}

\citet{1989A&A...209..327J} studied the mass loss in late-type
supergiants using three different damping mechanisms: nonlinear
damping, resonant surface damping, and turbulent damping. They
showed that these damping mechanisms have similar behaviors in the
wind. In Figure~\ref{fig.fluxo}, we compare the wave flux $\phi_A$
calculated by \citet{1989A&A...209..327J}, damped by the resonant
surface damping, and the frequency-integrated flux calculated by the
present model; that is, damped by the dust-cyclotron mechanism. It can be seen
that the damping of the waves by the dust-cyclotron mechanism is not
as efficient as any of the others mechanisms studied by
\citet{1989A&A...209..327J}: at $\sim 4~r_0$ the wave flux has droped
by more than a factor of $10^2$ in the model of
\citet{1989A&A...209..327J}. Also, the dust-cyclotron damping
mechanism is not as strong as the exponential decay that was simulated
in Paper I. Nevertheless, our model can reproduce observational data
such as the mass loss rate and the terminal velocity for the wind of a
typical K5 supergiant star.

\section{CONCLUSIONS}\label{sec.conc}
In this work, we consider an outward-directed flux of damped Alfv{\'e}n
waves as an acceleration mechanism for driving late-type stellar winds.
As these winds present great amounts of dust particles and as the dust
can modify the wave modes, we study here how the dust can affect the
propagation of Alfv{\'e}n waves, taking into account a
specific damping mechanism, dust-cyclotron damping, which can be
understood as follows. A charged grain immersed in a magnetized plasma
possesses a cyclotron frequency associated with it. If Alfv{\'e}n waves are
present in this plasma, their propagation near the dust-cyclotron
frequency can be affected. As different grain sizes carry different
charges, there will be a cyclotron frequency associated with each grain
size. In this sense, a dust size distribution has several
dust-cyclotron frequencies, which will affect a broad band of
Alfv{\'e}n wave frequencies. 

We also calculated the radiation pressure exerted on grains. We included
this force as another acceleration mechanism of the wind. As we
adopted great amouns of small grains, we observed only a negligible
difference in the wind velocity profile from what we would have
obtained if we did not consider radiation pressure.  

Our model also assumes a diverging geometry for the magnetic field
lines. This assumption is based on the observations of coronal holes
in the Sun, which present a superradial expansion of magnetic field
lines. Thus, the mass, momentum, and energy equations are obtained and
then solved in a self-consistent approach. 

Our results of wind velocity and temperature profiles for a
typical K5 supergiant star shows compatibility with observations. We
also show that, considering the presence of charged dust particles,
the wave flux is less damped due to dust-cyclotron damping than it
would be if we considered some other damping mechanisms studied in the
literature, such as nonlinear damping, resonant surface damping, and
turbulent damping.

\acknowledgments
The authors would like to thank the Brazilian agencies FAPESP (under
grant \mbox{02/10846-0}) and CNPq (under grant \mbox{304523/90-9}) for
financial support.

\clearpage

%\clearpage

\begin{table}
\begin{center}
\caption{Initial parameters adopted for the wind of a typical K5
  supergiant star.\label{parameters}} 
\begin{tabular}{ccl}
\tableline\tableline
Parameter & Value& Unity \\
\tableline
$r_0$ & $400$ &  R$_\odot$\\
$M_\star$ & 16 & M$_\odot$ \\
$\phi_{A_{0}}$ & $5.5 \times 10^6$ & erg cm$^{-2}$ s$^{-1}$  \\
$\rho_0$ & 1.07 $\times 10^{-13}$ & g cm$^{-3}$ \\ 
$u_0$ & $1.5 \times 10^5$ & cm s$^{-1}$  \\
$B_0$ & $10$ & G \\
$T_0$ & $3500$ & K \\
$S$ & $4.0$ & - \\ 
$r_t$ & $3.16$ & $r_0$ \\
\tableline
\end{tabular}
\end{center}
\end{table}

%\clearpage

\begin{table}
\begin{center}
\caption{Maximum and minimum calculated initial dust charges and cyclotron
  frequencies related to the minimum grain radius ($a_{\rm min} = 5.0
  \times 10^{-7}$~cm) and the maximum grains radius ($a_{\rm max} = 2.5
  \times 10^{-5}$~cm).\label{parameters_grains}}
\begin{tabular}{ccl}
\tableline\tableline
Parameter & Value& Unity \\
\tableline
$q_{\rm min}$ & $2.6$ & $e$\\
$q_{\rm max}$ & $131.2$ & $e$\\
$\omega_{\rm min}$ & $9.7 \times 10^{-5}$ & rad s$^{-1}$\\
$\omega_{\rm max}$ & $2.4 \times 10^{-1}$ & rad s$^{-1}$\\
\tableline
\end{tabular}
\end{center}
\end{table}


\begin{thebibliography}{}
\bibitem[Ayres et al.(2003)]{2003ApJ...598..610A}
Ayres, T. R., Brown, A., \& Harper, G. M 2003, \apj, 598, 610
\bibitem[Banerjee et al.(1998)]{1998A&A...339..208B}
{Banerjee}, D., {Teriaca}, L., {Doyle}, J.~G., {Wilhelm}, K. 1998,
\aap, 339, 208
\bibitem[Blackman et al.(2001)]{2001Natur.409..485B}
Blackman, E. G., Frank, A., Markiel, J. A., Thomas, J. H. \& Van Horn,
H. M. 2001, Nature, 409, 485
\bibitem[Carpenter et al.(1995)]{1995ApJ...444..424C}
{Carpenter}, K.~G., {Robinson}, R.~D., \& {Judge}, P.~G. 1995, \apj,
444, 424
\bibitem[Cassinelli(1979)]{1979ARA&A..17..275C}
Cassinelli, J. P. 1979, \araa, 17, 275
\bibitem[Cramer(2001)]{cramer2001}
Cramer, N.: 2001, in "The Physics of Alfv{\'e}n Waves", Wiley, Berlin
\bibitem[Cramer et al.(2002)]{2002PhPl....9.4845C}
Cramer, N., Verheest, F., \& Vladimirov, S. 2002, Phys. Plasmas, 9,
4845
\bibitem[Cranmer \& van Ballegooijen(2005)]{2005ApJS..156..265C}
Cranmer, S. R., \& van Ballegooijen, A. A. 2005, \apjs, 156, 265 
\bibitem[Dominik et al.(1989)]{1989A&A...223..227D}
Dominik, C., Sedlmayr, E., \& Gail, H. P. 1989, \aap, 223, 227
\bibitem[Dorch(2004)]{2004A&A...423.1101D}
Dorch, S. B. F. 2004, \aap, 423, 1101
\bibitem[Elitzur \& Ivezi{\'c}(2001)]{2001MNRAS.327..403E}
Elitzur, M., \& Ivezi{\'c}, Z. 2001, \mnras, 327, 403
\bibitem[Esser et al.(2005)]{2005ApJ...629L..61E}
Esser, R., Lie-Svendsen, \O., Janse, {\AA}.M., \& Killie, M. A. 2005,
\apjl, 629, 61
\bibitem[Falceta-Gon{\c c}alves \& Jatenco-Pereira(2002)]{2002ApJ...576..976F}
{Falceta-Gon{\c c}alves}, D. \& {Jatenco-Pereira}, V. 2002, \apj, 576, 976,
Paper I 
\bibitem[Falceta-Gon{\c c}alves et al.(2003)]{2003ApJ...597..970F}
Falceta-Gon{\c c}alves, D., de Juli, M. C., \& Jatenco-Pereira,
V. 2003, \apj, 597, 970 
\bibitem[Gilman(1974)]{1974ApJS...28..397G} 
Gilman, R.~C. 1974, \apjs, 28, 397
\bibitem[Goertz(1989)]{g89} 
Goertz, C. 1989, Rev. Geophys., 27, 271
\bibitem[Hartmann \& MacGregor(1980)]{1980ApJ...242..260H} 
Hartmann, L., \& MacGregor, K. B. 1980, \apj, 242, 260
\bibitem[Hartmann et al.(1982)]{1982ApJ...261..279H} 
Hartmann, L.,  Edwards, S., \& Avrett, E. 1982, \apj, 261, 279
\bibitem[Heras \& Hony(2005)]{2005A&A...439..171H}
Heras, A. M., \& Hony, S. 2005, \aap, 439, 171
\bibitem[H{\"o}fner et al.(2003)]{2003A&A...399..589H}
H{\"o}fner, S., Gautschy-Loidl, R., Aringer, B., \& J{\o}rgensen,
U. G. 2003, \aap, 399, 589
\bibitem[Hollweg(1973)]{1973ApJ...181..547H}
Hollweg, J. V. 1973, \apj, 181, 547
\bibitem[Holzer et al.(1983)]{1983ApJ...275..808H}
Holzer, T. E.,  Fl\aa\ , T., \& Leer, E. 1983, \apj, 275, 808
\bibitem[Jatenco-Pereira \& Opher(1989)]{1989A&A...209..327J}
Jatenco-Pereira, V., \& Opher, R. 1989 \aap, 209, 327
\bibitem[Knapp(1985)]{1985ApJ...293..273K}
Knapp, G.R. 1985, \apj, 293, 273
\bibitem[Kuin \& Hearn(1982)]{1982A&A...114..303K}
Kuin, N. P. M., \& Hearn, A. G. 1982, \aap, 114, 303
\bibitem[Lamers \& Cassinelli(1999)]{lamerscassinelli}
Lamers, H. J. G. L. M., \& Cassinelli, J. P. 1999, Introduction to
stellar winds (New York: Cambridge University Press)
\bibitem[Liberatore et al.(2001)]{2001A&A...377..522L}
Liberatore, S., Lafon, J.-P. J., \& Berruyer, N. 2001, \aap, 377, 522
\bibitem[Lie-Svendsen et al.(2002)]{2002ApJ...566..562L}
Lie-Svendsen, O., Hansteen, V. H., \& Leer, E. 2002, \apj, 566, 562
\bibitem[Mathis et al.(1977)]{mrn77}
Mathis, J., Rumpl, W., \& Nordsiek, K. 1977, \apj, 217, 425 
\bibitem[Mendis \& Rosenberg(1992)]{mr92}
Mendis, D., \& Rosenberg, M. 1992, IEEE Trans. Plasma. Sci., 20, 929
\bibitem[Mendis \& Rosenberg(1994)]{mr94}
Mendis, D., \& Rosenberg, M. 1994, \araa, 32, 419
\bibitem[Parker(1965)]{parker}
Parker, E. (1965), Space Science Reviews, 4, 666
\bibitem[Peter \& Judge(1999)]{1999ApJ...522.1148P}
Peter, H., \& Judge, P. G. 1999, \apj, 522, 1148
\bibitem[Pillip et al.(1987)]{1987ApJ...314..341P}
Pillip, W., Morfill, G., Hartquist, T., \& Havnes, O. 1987, \apj, 314, 341
\bibitem[Rodgers \& Glassgold(1991)]{1991ApJ...382..606R}
Rodgers, B., \& Glassgold, A. E. 1991, \apj, 382, 606
\bibitem[Rosenthal et. al(2002)]{rosenthal02}
Rosenthal, C. S., Bogdan, T. J., Carlsson, M., Dorch, S. B. F.,
Hansteen, V., McIntosh, S. W., McMurry, A., Nordlund, \AA, \& Stein,
R. F. 2002, \apj, 564, 508
\bibitem[Sandin \& H{\"o}fner(2003)]{2003A&A...404..789S}
Sandin, C., H{\"o}fner, S. 2003, \aap, 404, 789
\bibitem[Schmutzler \& Tscharnuter(1993)]{1993A&A...273..318S} 
Schmutzler, T., \& Tscharnuter, W. M. 1993, \aap, 273, 318
\bibitem[Shukla(1992)]{1992PhyS...45..504S}
Shukla, P. 1992, Phys. Scr., 45, 504
\bibitem[Soker \& Zoabi(2002)]{2002MNRAS.329..204S}
Soker, N., \& Zoabi, E. 2002, \mnras, 329, 204
\bibitem[Suzuki \& Inutsuka(2005)]{suzuki05}
Suzuki, T. K., \& Inutsuka, S. 2005, \apjl, 632, 52
\bibitem[Tu et al.(1989)]{1989JGR....9411739T}
Tu, C. Y.,  Marsch, E., \& Thieme, K. M. 1989, \jgr, 94, 11739
\bibitem[Tu et al.(2005)]{2005Sci...308..519T}
Tu, C.-Y., Zhou, C., Marsch, E., Xia, L.-D., Zhao, L., Wang, J.-X.,
Wilhelm, K. 2005, Science, 308, 519
\bibitem[Vladimirov(1997)]{1997Ap&SS.256...85V}
Vladimirov, S. V. 1997, \apss, 256, 85
\bibitem[Vlemmings et al.(2002)]{2002A&A...394..589V}
Vlemmings, W. H. T., Diamond, P. J., \& van Langevelde, H. J. 2002, \aap,
394, 589
\bibitem[Wardle \& Ng(1999)]{wn99}
Wardle, M., \& Ng, C. 1999, \mnras, 303, 239
\bibitem[Woitke \& Niccolini(2005)]{2005A&A...433.1101W}
Woitke, P., \& Niccolini, G. 2005, \aap, 433, 1101

\end{thebibliography}
\end{document}